\documentclass[a4paper,preprint,12pt,authoryear,review]{elsarticle}

\usepackage[gen]{eurosym}
\usepackage[hidelinks]{hyperref}
\usepackage{placeins}

\usepackage{amssymb}
\usepackage{amsmath}
\usepackage{layaureo}
\usepackage{booktabs}
\usepackage{tabularx}
\usepackage{makecell}
\usepackage{subcaption}
\usepackage{pifont}
\usepackage{tabularx}
\usepackage{tablefootnote}
\usepackage{setspace}
\usepackage{comment}
\usepackage{tikz}
\usepackage{multirow}
\usepackage[T1]{fontenc}
\usepackage{threeparttable}
\journal{arXiv}

\begin{document}
\begin{frontmatter}

\title{Most certainly certain? The Impact of Contract for Difference Design on Renewables' Strike Prices and Electricity Market Risks
\tnoteref{label1}}
\author[1]{Silke Johanndeiter\corref{cor1}}
\cortext[cor1]{silke.johanndeiter@rub.de}
\affiliation[1]{organization={Ruhr-Universität Bochum},
    addressline={Universitätsstr. 150}, 
    city={44801 Bochum},
    country={Germany}}

\author[1]{Jonas Finke}

\author[2]{Justus Heuer}
\affiliation[2]{organization={Universität zu Köln},
    addressline={Albertus-Magnus-Platz}, 
    city={50923 Köln},
    country={Germany}}


\begin{singlespace}
\begin{highlights}
\item Derivation of optimal strike prices in competitive CfD auction under uncertainty 
\item Investor expectations formed through 36 market scenarios in energy system model
\item CfDs significantly reduce both profit and consumer price volatility
\item CfD design affects profit volatility, but hardly consumer price volatility
\item Reference power plant drives both profit volatility and system-friendly invest
\end{highlights}

\begin{abstract}
Weather, technological and regulatory uncertainties expose actors in highly renewable electricity markets to substantial price and volume risks. Two-way Contracts for Difference (CfDs) can mitigate these risks. They stipulate payments between the government and generators of renewable electricity based on the difference of a strike and a reference price, whose definition and unit of payment differ between CfD designs. We study the effect of three different CfD designs on wind power profit and consumer price volatility under the consideration of uncertain market outcomes in a highly renewable, sector-coupled electricity market. First, we analytically derive optimal strike prices under uncertainty. Second, we numerically determine optimal strike prices based on market expectations retrieved from optimising a set of 36 market scenarios in an energy system model. Third, we study the distribution of ex post market revenues, CfD payments and consumer prices across all 36 scenarios. Compared to purely market-based consumer prices and investor profits, we find all CfDs to significantly reduce volatility. For consumer prices, results show no substantial differences between CfD designs. For investor profits, we identify the highest volatility reduction under a capacity-based CfD with a reference price similar to power plants' individual market revenues. Since such a CfD design is known to diminish the effect of price signals on investment decisions, our results reveal a trade-off between incentivising system-friendliness and reducing investor risk.
\end{abstract}

\end{singlespace}
\end{frontmatter}

\newpage
\section{Introduction}
\label{sec:intro}
The transition to a climate-neutral energy system exposes electricity market actors to significant risks. Consequences of a severe price shock in the European electricity market became evident during the energy crisis in 2022, when renewables achieved windfall profits at the expense of rising consumer bills \citep{anton2023}. As a response, the European Union introduced two-way Contracts for Difference (CfDs) as the mandatory support scheme for renewables due to their capability to mitigate risks for both generators and consumers of electricity \citep{eu2024}. They stipulate payments between the government and producers of renewable electricity defined by the difference of an ex post realised reference price and an ex ante agreed-upon strike price over a given contracting period -- typically the expected plant operation time \citep{schlecht2024}. Two-way CfDs generally decrease generator revenues in times of high electricity prices, and increase them, when electricity prices are low, affecting both average revenues and revenue volatility. If CfD expenses and revenues are passed on to consumers via levies, they impact consumer prices in the opposite direction. There exist numerous design options of CfDs that affect their potential to reduce volatility of investor profits and consumer prices. Our work assesses the impact of three two-way CfD design options on these values in a highly renewable European electricity market.

\subsection{Derisking and financing costs}
In the last decade, renewables became competitive with fossil fuels in terms of levelised cost of electricity (LCOE) \citep{kost2024}. Since then, the primary objective of renewable support schemes has shifted from increasing average revenues to reducing revenue uncertainty \citep{beiter2024}. Underlying revenue risks can be split into two components \citep{schlecht2024}:
\begin{enumerate}
    \item Volume risk: The uncertainty about the amount of generated electricity, which mainly depends on the weather and curtailment.
    \item Price risk: The uncertainty about the market value of the generated electricity, which depends on overall market trends, the total fleet of generators, infrastructure constraints, and political factors. 
\end{enumerate}

Several studies provide empirical or model-based evidence that lowering these revenue risks leads to reduced financing costs for renewable investments. For the Australian electricity market, \citet{gohdesetal2022} show an empirical correlation between the availability of long-term risk hedging options for renewable investments and lower LCOEs. Using a liquidity management model, \citet{dukan2025} demonstrate how revenue stabilisation with CfDs decreases financing costs by allowing to decrease debt size. In a costly state verification framework, \citet{soysal2025} predicts the effect of revenue variability on the default risk and, subsequently, cost of capital. \citet{leveque2026} combine a model-based calculation of cost of debt and equity with global data on policy regimes and cost of capital across different countries to show an empirical correlation between risk reduction through policy and lower financing costs. These, in turn, have a significant positive effect on the expansion of renewable electricity generation as a capital-cost-intensive technology \citep{bachneretal2019,eglietal2020} and thus also reduce total costs in highly renewable power system \citep{helistoeetal2024}. 

There exist private markets for long-term contracts to hedge risks. Forward and future contracts provide hedges up to 5 or even 10 years ahead of delivery \citep{eex2025}. In the context of renewables, the market for private power purchase agreements (PPAs) with contract durations of 15 years or more is growing \citep{ppa2025}. However, the market for long-term private-to-private risk hedging instruments still lacks liquidity -- an issue coined the missing risk market problem by \cite{dimanchevetal2024}. Various instruments, such as market-maker obligations, can increase market liquidity \citep{schittekatte2023}. Nevertheless, some scholars argue for the necessity of governmentally issued long-term contracts \citep{fabra2022,neuhoff2023}. In addition to addressing the lack of liquidity, they reduce price risks for household consumers, who lack access to long-term private risk-hedging instruments, such as PPAs, whose offtakers are dominated by corporate consumers \citep{Mili2025}. 

\subsection{CfD design elements}
\label{sec:intro_design}
Two-way CfDs are such governmentally issued long-term contracts.
They have two main design elements that affect their potential to reduce risk and their impact on system costs. The first element is the \textit{reference price}. It is mainly determined by the length of the reference period, i.e., the time frame used to calculate it, and the selection of reference generators, i.e., whether all or only some power plants are considered. The simplest CfD types compare individual market performance in (quarter-)hourly trading windows to the strike price. Such simple CfDs entirely hedge price risks \citep{wagner2026}. However, this also implies that they eliminate market price signals, thereby incentivising to design, site and operate a power plant in a manner that merely maximises full load hours. Conversely, more sophisticated CfDs preserve market price signals in both the investment and operational phase by decoupling individual market revenues from CfD payments. These CfD types use market revenues of a reference power plant (e.g., the market's average) over longer periods (e.g., a month or a year) as reference price. In this case, power plants can retain market revenues above the reference price. Thus, this CfD design incentivises investments in power plants with an above-average market performance \citep{schmidt2013,may2017,meus2021}. These power plants can be deemed more \textit{system-friendly}, as higher market values typically imply a higher temporal match of a power plant's supply and load, therefore inducing lower balancing costs \citep{ueckerdt2013}. Similarly, \cite{diallo2024} associate market values with the social value of a power plant. For highly renewable energy systems, CfDs that retain market price signals are indeed shown to lower system costs \citep{veenstra2024,johanndeiter2025}. However, such CfD designs also imply that not all price risks are fully hedged \citep{wagner2026}. 

The second important design element of CfDs is the \textit{unit of payment}: Reference and strike prices can either be production- or capacity-based, influencing both risk reduction as well as investment and operational incentives. Production-based CfDs only capture price risks, while capacity-based CfDs also hedge against volume risk. Furthermore, production-based CfDs are known to distort dispatch decisions in the sense that they cause power plants to be dispatched at prices below their marginal production costs \citep{devos2015,andor2016,frondel2022}. If the reference price is equal to the (quarter-)hourly spot market price, they incentivise to maximise output regardless of the electricity price. If the reference period is longer, they induce so-called virtual variable costs or revenues equal to anticipated CfD payments to or from the government. Virtual costs cause power plants to be dispatched only at price levels above the amount of the CfD payment, whereas virtual revenues lead to dispatch even at negative prices \citep{rosnes2014,pahle2016,chaves2017,frey2020}. The cited studies are described in more detail in \cite{johanndeiter2025}, while conceptual discussions of these and further design aspects are provided by \citet{huntington2017},  \citet{newbery2023} and \citet{schlecht2024}. 

\subsection{Market risks in the energy transition}
Our work focuses on the impact of CfD design on risks in highly renewable electricity markets. A main driver of both volume and price risks in an energy system dominated by weather-dependent power supply is weather risk \citep{blazquez2018,alonzo2022,jimenez2024}. Volume risk is additionally affected by curtailment risk \citep{egli2020}. Further risks arise from technological uncertainty that, amongst others, is reflected in unknown cost and demand levels: Cost degression rates of both renewables and battery storages have been systematically underestimated \citep{xiao2021,ziegler2021}. Electricity demand levels for Europe can vary by 2000 TWh in just one single study \citep{eucom2018}, and future hydrogen demand levels are even more uncertain \citep{hanley2018}. Furthermore, fuel costs are shown to continue to affect market prices in highly renewable electricity markets \citep{ruhnau2022,johanndeiter2025b} and historically proven to be volatile \citep{furio2012}. Furthermore, regulatory uncertainty exposes investors to significant price and volume risks, including uncertainty regarding the introduction of capacity remuneration mechanisms, coal or nuclear phase outs, renewable targets, revenue caps or CO$_2$ prices. Together with other macro risks, such as general macroeconomic and political stability, \cite{leveque2026} find this type of risk to constitute the main share of renewable projects' cost of capital.
\subsection{CfD design and derisking}
In a CfD setting, the above-described uncertainties are reflected in CfDs' underlying strike prices that are agreed upon prior to the investment and kept constant for the entire contract duration. Compliant with European regulation \citep{cep2018}, strike prices are determined within competitive auctions that allocate CfDs to specific renewable energy capacities \citep{anatolitis2017}.  
\citet{diallo2024} and \citet{johanndeiter2025} show how optimal strike prices in a competitive auction for a CfD take into account the above-described incentives: If the CfD auction for the strike price is perfectly competitive, an investor bids a strike price that sets expected profits including CfD payments over the economic lifetime of the investment project to zero. Therefore, the strike price also reflects the relative market performance of a power plant, if the CfD design includes incentives for system-friendly investments. 
However, expected electricity generation and market revenues over economic lifetimes of typically 15 or more years are subject to the above-described uncertainties.
Therefore, investors are required to form expectations on these revenue determinants to calculate their strike price bids for the auction. Actual market revenues and subsequent CfD payments, in turn, are determined based on ex post outcomes, i.e., market prices and electricity generation. Hence, ex post investor profits are not necessarily zero, if expected market outcomes used to determine strike prices deviate from ex post outcomes.

Previous work analyses the effect of support scheme design on risk exposure, financing cost and/or strike prices: Using models of renewable projects' financing structure, \cite{neuhoffetal2022} and \cite{dukan2023} show that
strike prices are lower under a one-way (also deemed sliding market premia) than under a two-way CfD, while financing costs are higher due to their greater exposure to electricity market price risks. \cite{breitschopf2022} provide empirical evidence that lower market price exposure indeed decreases financing costs. Relatedly, in a real-options approach applied to a Nordic wind power case study, \cite{boomsma2012} conclude that renewable support schemes with a higher risk exposure, particularly tradable green certificates compared to feed-in tariffs, require higher revenues to trigger investments. \cite{boomsma2015}, in turn, show how risk exposure can be decreased by diversification and \cite{fagiani2013} highlight the impact of risk preferences on the cost-efficiency of the two support schemes. Based on a conceptual discussion, \cite{hiroux2010} suggest that price premia strike an ideal balance between support schemes that expose power plants to accurate market price signals and risk reduction. Finally, \cite{wagner2026} study the interaction of forward markets and CfDs and show that CfD designs with shorter reference periods decrease forward market liquidity -- possibly undermining the purpose of CfDs. Yet, besides \cite{wagner2026}, existing studies are limited to simple support scheme designs, such as feed-in tariffs, fixed market premia or sliding market premia/CfDs based on the hourly spot market price. Furthermore, to the best of our knowledge, there exists no previous work that focuses on the deviation of ex post market realisations from ex ante expectations used to determine a support scheme's underlying strike price. \\

We close this gap by analysing profit volatility of wind onshore power plants and consumer prices under three designs of two-way CfDs. The novelty of our work lies in the consideration of uncertain market outcomes in a sector-coupled, highly renewable European power system that are reflected in the CfDs' underlying strike prices. Furthermore, we uniquely consider CfD designs that vary in terms of the reference price and the unit of payment. Particularly, we consider two production-based CfDs and one capacity-based CfD. One production-based CfD's reference price is defined as the hourly spot market price, while the other considered CfD types use average revenues of all wind power plants in one bidding zone as reference price. Furthermore, we consider two wind production profiles by bidding zone: one with above-average full load hours and one with an above-average market value, where the latter would be considered more \textit{system-friendly} according to our definition above. 
In this study, we first build an analytic model to determine strike price bids for three different CfD designs based on ex ante price and volume expectations (Section \ref{sec:theory}). Second, we retrieve required expectations from a sector-coupled power system model that we apply to optimise a set of scenarios of a highly renewable European electricity market that reflect both weather risks and technological risks (Section \ref{sec:method}). Third, we use the optimisation results to determine ex ante optimal strike prices by CfD design and calculate each power plant's revenues from market sales, CfD payments and costs that would result ex post in each scenario realisation. This allows us to determine the level of cost recovery as well as subsequent effects on consumer prices, if CfD payments are financed via a levy. We then analyse the volatility of cost recovery and consumer prices by CfD type across scenarios (Section \ref{sec:results}). Finally, Section \ref{sec:conclusion} concludes.

\section{Theoretical background}
\label{sec:theory}
Consider a renewable power plant $i \in I$ located in bidding zone $n \in N$.
Its annual profit
\begin{align}
\label{eq:profit}
  \Pi^\text{type}_{i,n}=\sum_{t\in T}q_{t,i,n}p_{t,n}-C_{i,n}+P^\text{type}_{i,n}  
\end{align}
is given by spot market revenues minus costs plus net CfD payments from or to the government.
Spot market revenues result from selling power quantities $q_{t,i,n}$ for prices $p_{t,n}$ in every time step $t \in T$ of the year.
Costs
\begin{align}
    \label{eq:cost}
    C_{i,n}= c \cdot \sum_{t \in T} q_{t,i,n}+A \cdot M \cdot Q_{i,n}    
\end{align}
include variable costs $c$ incurred per quantity produced and investment costs $M$ incurred per installed capacity $Q_{i,n}$ that are annualised with annuity factor $A$.
CfD payments and therefore profits $P^\text{type}_{i,n}$ differ by type $\in \{ \text{basic, 2way, fin}\}$ of CfD.
They depend on a possibly time-dependent reference price $p_{t,n}^{R,\text{type}}$, which is uniform within each bidding zone and determined ex post, as well as an individual strike price $S_{i,n}^\text{type}$ per power plant that is set ex ante.
For simplicity, we assume that $c$, $A$ and $M$ do not depend on bidding zone or power plant, i.e.\ we consider only one power plant technology with homogeneous cost assumptions across countries. Hence, power plants $i \in I$ only differ in terms of their generation profiles.

In the following, we introduce the three CfD types together with how strike prices are formed.
Then we derive optimal strike prices for each CfD type, first without and then with uncertainty.

\begin{table}[]
    \centering
    \footnotesize
    \caption{Overview of considered CfD designs}
 \begin{tabular}{c c c}
         \toprule
 CfD type & Unit of payment & Reference price  \\
\midrule Basic & \euro{}/MWh & Hourly spot market price \\
2way & \euro{}/MWh & Average wind market value  \\
Financial & \euro{}/MW & Average wind revenues per capacity \\
         \bottomrule
    \end{tabular}
 \label{tab:cfd_types}
\end{table}

\subsection{Definition of CfD types}

Following \citet{johanndeiter2025}, we consider two production-based CfDs, the basic and the 2way CfD that differ in terms of their reference price.
For the basic CfD, the reference price is the hourly market price, $p^\text{R,basic}_{t,n}=p_{t,n}$, so that payments are given by
\begin{align}
    \label{eq:payment_basic}
    P^\text{basic}_{i,n}=\sum_{t\in T}(S_{i,n}^\text{basic}-p_{t,n}) \, q_{t,i,n}.
\end{align}
For the 2way CfD, the reference price equals the yearly zonal market value, $p^\text{R,2way}_{t,n} = v_n$ for all $t \in T$, which is defined as the generation-weighted market revenues averaged over the entire year and all power plants in bidding zone $n$\footnote{Although all of the considered CfD designs are two-way CfDs, we name this type of CfD ``2way'' as it most closely corresponds to two-way CfDs as implemented in practice. In contrast, the ``basic'' CfD is equivalent to a feed-in tariff.}. That is,
\begin{align} \label{eq:payment_2way}
    P^\text{2way}_{i,n}= (S_{i,n}^\text{2way}-v_n) \sum_{t \in T}q_{t,i,n},
\end{align}
where
\begin{align}
    \label{eq:v_n}
    v_n = \frac{\sum_{t \in T}\sum_{i \in I}q_{t,i,n}\,p_{t,n}}{\sum_{t\in T}\sum_{i \in I}q_{t,i,n}}.
\end{align}

Substituting the CfD payments (\ref{eq:payment_basic}) and (\ref{eq:payment_2way}) in the profit function (\ref{eq:profit}) shows that in some cases, generation may increase profit even at negative prices or decrease profit even at prices above the variable cost. Therefore, both production-based CfD types may distort optimal dispatch.

Moreover, we consider the capacity-based \emph{financial} CfD as proposed by \citet{schlecht2024} that resembles the capacity-based support scheme proposed by \citet{huntington2017}.
For this CfD type, the reference price (now in units of \euro/MW) is given by the average annual revenues per unit capacity of a reference power plant, $p^\text{R,fin}_{t,n} = r_n$.
Here, we form the average across all power plants in one bidding zone, i.e.
\begin{align}
    r_n=\frac{\sum_{t \in T}\sum_{i \in I}q_{t,i,n}\,p_{t,n}}{\sum_{i \in I}Q_{i,n}},
\end{align}
so that CfD payments are given by
\begin{align}
    \label{eq:payment_fin}
    P^\text{fin}_{i,n}=Q_{i,n}(S_{i,n}^\text{fin}-r^\text{fin}_n).
\end{align}

\subsection{Optimal strike prices without uncertainty}
Compliant with European law \citep{cep2018}, we assume CfDs to be awarded within a tendering procedure. Particularly, we suppose a limited number of renewable capacities to be awarded within a competitive pay-as-bid auction for the strike price for a contract duration that equals power plants' economic lifetime.

Then, a rational, risk-neutral investor chooses a strike price that sets expected profits over lifetime to zero.
For this case, \citet{johanndeiter2025} derive optimal ex ante strike prices $S_{i,n}^\text{type*}$ under perfect foresight.

For the basic CfD, the optimal strike price equals the individual levelised cost of electricity (LCOE),
\begin{align}
{S_{i,n}^\text{basic*}}= \frac{C_{i,n}}{\sum_{t \in T}q_{t,i,n}}\equiv\text{lcoe}_{i,n}.
\label{eq:sbasic}
\end{align}
For the 2way CfD, the optimal strike price equals the LCOE, but adopted for the difference between the power plant-averaged market value $v_n$ and the individual market value $v_{i,n}$,
\begin{align}
\label{eq:s2way}
S_{i,n}^\text{2way*} &=\frac{C_{i,n}}{\sum_{t\in T}q_{t,i,n}} +\frac{\sum_{t \in T}\sum_{i\in I}q_{t,i,n}p_{t,n}}{\sum_{t \in T}\sum_{i\in I}q_{t,i,n}}-\underbrace{\frac{\sum_{t \in T}q_{t,i,n}p_{t,n}}{\sum_{t \in T}q_{t,i,n}}}_{\equiv v_{i,n}} \notag \\  &=\text{lcoe}_{i,n}+v_{n}-v_{i,n}.
\end{align}
For the financial CfD, the optimal strike price (in \euro/MW) follows the same pattern, but costs and revenues are normalised per capacity instead of generation.
That is, with market revenues per capacity given by $r_{i,n}$ for an individual power plant and by $r_n$ when averaged over all power plants, the optimal strike price equals
\begin{align}
\label{eq:sfin}
{S_{i,n}^\text{fin*}=\frac{C_{i,n}}{Q_{i,n}}+\underbrace{\frac{\sum_{t \in T}\sum_{i\in I}q_{t,i,n}p_{t,n}}{\sum_{i\in I}Q_{i,n}}}_{\equiv r_n} - \underbrace{\frac{\sum_{t \in T}q_{t,i,n}p_{t,n}}{Q_{i,n}}}_{\equiv r_{i,n}}
=\frac{C_{i,n}}{Q_{i,n}}+r_n-r_{i,n}}.
\end{align}

Here, we assume that the total generation $\sum_{t \in T}q_{t,i,n}$ and the capacity $Q_{i,n}$ are strictly positive. The definitions of $\text{lcoe}_{i,n}, v_n, v_{i,n}, r_n, r_{i,n}$ can simply be extended to equal zero otherwise.

The optimal strike prices illustrate investment incentives set by each type of CfD: in a competitive auction, the basic CfD is awarded to power plants with the lowest LCOE, typically driven by a high expected number of full load hours. Conversely, the 2way and financial CfDs allow power plants to accept a strike price lower (higher) than their expected average costs, if expected market revenues are higher (lower) than the reference price. Thus, they incentivise system-friendly investment decisions as indicated by a higher market value (cf.\ Section \ref{sec:intro_design}).


\subsection{Optimal strike prices under uncertainty}
\label{sec:theory_sp_unc}
Several variables in Equations (\ref{eq:sbasic}), (\ref{eq:s2way}) and (\ref{eq:sfin}) are uncertain at the time of the auction, requiring investors to form expectations. We assume that the uniform cost parameters $c$, $A$ and $M$ feeding into $C_{i,n}$ are known to investors with certainty.
Hence, for calculating its optimal strike price, the power plant $i$ in bidding zone $n$ needs the following parameters for the own bidding zone that remain uncertain:
\begin{itemize}
    \item Hourly market prices $p_{t,n}$ for all $t\in T$ 
    \item Hourly capacity factor $f_{t,i,n} \equiv q_{t,i,n}/Q_{i,n}$ of the own power plant for all $\,t\in T$
    \item Hourly capacity factors $f_{j,t,n}$ of all other power plants $j \in I \setminus \{i\}$ for all $t\in T$
    \item Ratio of the capacity of power plant $i$ to total capacities (of the considered technology) $w_{i,n} \equiv Q_{i,n} / \sum_{i \in I} Q_{i,n}$ for all $i\in I$.
\end{itemize}

We assume investors to form expectations as expected values $\mathbb{E}_{\{s\in S\}}[\cdot]$ of the relevant uncertain terms throughout a discrete, finite set of electricity market scenarios represented from here on by introducing a new index $s \in S$.
Accordingly, optimal strike prices under uncertainty $\Tilde{S}_{i,n}^\text{type}$ can be derived by substituting the cost Equation (\ref{eq:cost}) and $\sum_{t \in T} f_{t,i,n,s} = \sum_{t\in T} q_{t,i,n,s} / Q_{i,n,s} $ as well as using linearity of expectation in Equations (\ref{eq:sbasic}), (\ref{eq:s2way}) and (\ref{eq:sfin}).

For the basic CfD, uncertainty only occurs in the own capacity factor, 
\begin{align}
    \label{eq:sbasic_unc}
    \Tilde{S}_{i,n}^\text{basic} &= \mathbb{E}_{\{s \in S\}}[\text{lcoe}_{i,n,s}] = c + \mathbb{E}_{\{ s \in S\}} \left[ \frac{A \cdot M}{\sum_{t\in T} f_{t,i,n,s}} \right] \notag \\
    &= c +\frac{A \cdot M}{\sum_{t \in T}\mathbb{E}_{\{s \in S\}}[f_{t,i,n,s}]}.
\end{align}
For the 2way CfD, we approximate 
\begin{align}
    \label{eq:s2way_unc}
    \Tilde{S}_{i,n}^\text{2way} 
    &= \mathbb{E}_{\{s \in S\}}[\text{\text{lcoe}}_{i,n,s}] + \mathbb{E}_{\{s \in S\}}[v_{n,s}] - \mathbb{E}_{\{s \in S\}}[v_{i,n,s}] \notag \\
    &= c +\frac{A \cdot M}{\sum_{t=1}^{T}\mathbb{E}_{\{s \in S\}}[f_{t,i,n,s}]}
   + \mathbb{E}_{\{s \in S\}} \left[\frac{\sum_{t=1}^{T} \sum_{i\in I} 
           [f_{t,i,n,s} \cdot p_{t,n,s}]}
          {\sum_{t=1}^{T} \sum_{i \in I} 
          [f_{t,i,n,s}]} \right] \notag \\
    &\quad - \mathbb{E}_{\{s \in S\}} \left[\frac{\sum_{t=1}^{T} 
            [f_{t,i,n,s} \cdot p_{t,n,s}]}
           {\sum_{t=1}^{T} 
           [f_{t,i,n,s}]} \right] \notag \\
           & \approx c +\frac{A \cdot M}{\sum_{t=1}^{T}\mathbb{E}_{\{s \in S\}}[f_{t,i,n,s}]}
   +\frac{\sum_{t=1}^{T} \sum_{i\in I} 
           \mathbb{E}_{\{s \in S\}} [f_{t,i,n,s} \cdot p_{t,n,s}]}
          {\sum_{t=1}^{T} \sum_{i \in I} 
          \mathbb{E}_{\{s \in S\}} [f_{t,i,n,s}]} \notag \\
    &\quad - \frac{\sum_{t=1}^{T} 
            \mathbb{E}_{\{s \in S\}} [f_{t,i,n,s} \cdot p_{t,n,s}]}
           {\sum_{t=1}^{T} 
           \mathbb{E}_{\{s \in S\}} [f_{t,i,n,s}]}
\end{align}        

by using a second order Taylor expansion $E\left(\frac{X}{Y}\right) \approx \frac{E(X)}{E(Y)} - \frac{\text{Cov}(X,Y)}{(E(Y))^2} + \frac{E(X)\text{Var}(Y)}{(E(Y))^3} \approx \frac{E(X)}{E(Y)}$ as in our case $\frac{\text{Cov}(X,Y)}{(E(Y))^2} + \frac{E(X)\text{Var}(Y)}{(E(Y))^3} \approx 0 $ \citep{stuart2010}.

For the financial CfD, we derive
\begin{align}
\label{eq:sfin_unc}
\Tilde{S}_{i,n}^\text{fin} 
&= \mathbb{E}_{\{s \in S\}}\left[\frac{C_{i,n,s}}{Q_{i,n,s}}\right]+\mathbb{E}_{\{s \in S\}}[r_n]-\mathbb{E}_{\{s \in S\}}[r_{i,n}] \notag \\
& = \sum_{t \in T}c\cdot\mathbb{E}_{\{s \in S\}}[f_{t,i,n,s}]+A \cdot M
   +  \sum_{i \in I}\sum_{t \in T} \mathbb{E}_{\{s \in S\}}\left[w_{i,n,s} \cdot f_{t,i,n,s} \cdot p_{t,n,s} \right] 
     \notag \\
& -         \sum_{t \in T}
        \mathbb{E}_{\{s \in S\}}[f_{t,i,n,s} \cdot p_{t,n,s}].
\end{align}

Now, we can compute the remaining expected values in Equations \eqref{eq:sbasic_unc}, \eqref{eq:s2way_unc} and \eqref{eq:sfin_unc} by assigning an equal weight of $\frac{1}{S}$ to each scenario, and using the covariance $\mathrm{Cov}\left[X,Y\right] = \mathbb{E} \left[ \left( X -\mathbb{E}\left[ X \right] \right) \left( Y - \mathbb{E}\left[Y\right] \right) \right] = \mathbb{E} \left[ X Y \right] - \mathbb{E} [X] \mathbb{E} [Y]$.

For the simple expected values we use
\begin{align}
\mathbb{E}_{\{s \in S\}}[f_{i,t,n,s}] = \frac{1}{S} \sum_{s=1}^S f_{t,i,n,s}
\end{align}
and
\begin{align}
\mathbb{E}_{\{s \in S\}}[p_{t,n,s}] = \frac{1}{S} \sum_{s=1}^S p_{t,n,s}
\end{align}
for products of two uncertain variables,
\begin{align}
\mathbb{E}_{\{s \in S\}}[f_{t,i,n,s} \cdot p_{t,n,s}] = 
        \mathbb{E}_{\{s \in S\}}[f_{t,i,n,s}] \cdot 
               \mathbb{E}_{\{s \in S\}}[p_{t,n,s}] 
           + \mathrm{Cov}_{\{s \in S\}}[f_{t,i,n,s}, p_{t,n,s}],
\end{align}
where the covariance for a finite, discrete set of scenarios can be simplified as
\begin{align}
\mathrm{Cov}_{\{s \in S\}}[f_{t,i,n,s}, p_{t,n,s}] =
\frac{1}{S} \sum_{s=1}^S (f_{t,i,n,s}-\mathbb{E}_{\{s \in S\}}[f_{t,i,n,s}])(p_{t,n,s}-\mathbb{E}_{\{s \in S\}}[p_{t,n,s}]).
\end{align}
Finally, this covariance can also be used to simplify the product of three variables 
\begin{align}
\mathbb{E}_{\{s \in S\}}[w_{i,n,s} \cdot f_{t,i,n,s} \cdot p_{t,n,s}] &= \mathbb{E}_{\{s \in S\}}[w_{i,n,s}] \cdot 
        \mathbb{E}_{\{s \in S\}}[f_{t,i,n,s}] \cdot 
               \mathbb{E}_{\{s \in S\}}[p_{t,n,s}] \notag \\
& + \mathbb{E}_{\{s \in S\}}[w_{i,n,s}] \cdot \mathrm{Cov}_{\{s \in S\}}[f_{t,i,n,s}, p_{t,n,s}] \notag \\ &+ \mathrm{Cov}_{\{s \in S\}}[w_{i,n,s},f_{t,i,n,s}\cdot p_{t,n,s}],
\end{align}
where the last covariance term may be dropped if assuming that the share of own capacity compared to total capacity is independent of the own revenues per unit capacity.

\newpage
\section{Methodology}
\label{sec:method}

We study volatility of wind power plant revenues and consumer costs under different types of CfDs in three steps.
First, we optimise an ensemble of deterministic scenarios of a fully decarbonised European electricity market with a high share of variable renewables.
Second, we assume the optimisation results to represent a set of plausible futures of the electricity market considered by investors. This allows us to determine optimal strike prices under uncertainty per CfD type in three bidding zones for two wind power plants each, based on the Equations in Section \ref{sec:theory_sp_unc}.
Third, we calculate ex post CfD payments resulting in each scenario realisation to study ex post cost recovery and consumer prices.

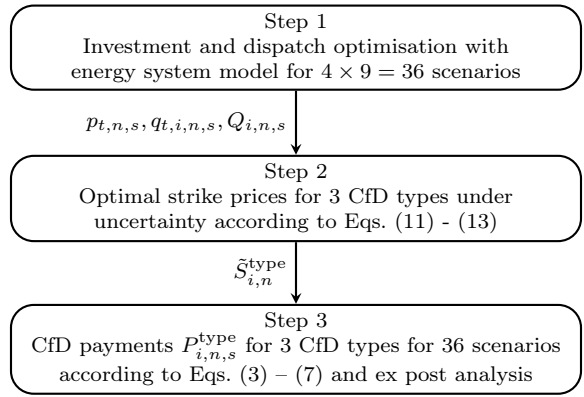
\begin{figure}[h!]
    \centering
    \scriptsize
    \begin{tikzpicture}[node distance=0cm,
                    agent/.style={
                    rectangle,minimum width=7.5cm,minimum height=1cm,rounded corners=3mm, thick,draw=black
                    },
                    arrow/.style={
                    thick,->,>=stealth
                    }
                    ]

    \node (Step2) [agent, align=center] {Step 2\\Optimal strike prices for 3 CfD types under\\uncertainty according to Eqs.\ (\ref{eq:sbasic_unc}) - (\ref{eq:sfin_unc})};
    \node (Step1) [agent, above of=Step2, yshift=2cm, align=center] {Step 1\\Investment and dispatch optimisation with\\energy system model for $4 \times 9 = 36$ scenarios};
    \node (Step3) [agent, below of=Step2, yshift=-2cm, align=center] {Step 3\\CfD payments $P^\text{type}_{i,n,s}$ for 3 CfD types for 36 scenarios\\according to Eqs.\ (\ref{eq:payment_basic}) -- (\ref{eq:payment_fin}) and ex post analysis};

    \draw [arrow] (Step1) -- node[anchor=east] {$p_{t,n,s},q_{t,i,n,s},Q_{i,n,s}$} (Step2);
    \draw [arrow] (Step2) -- node[anchor=east] {$\Tilde{S}_{i,n}^\text{type}$} (Step3);
    
    \end{tikzpicture}
    \caption{Methodology.}
    \label{fig:methodology}
\end{figure}


\subsection{Invest and dispatch optimisation for ensemble of scenarios}
\label{sec:esm}
We apply the sector-coupled European electricity market model developed by \cite{johanndeiter2025b} to optimise an ensemble of fully decarbonised market scenarios. The model is implemented in the energy system modelling and optimisation tool Backbone \citep{helistoe2019} and covers all EU-27 countries except for Malta, but including Great Britain, Switzerland and Norway, aggregated to 19 bidding zones -- all of which are connected by power and H$_2$ transmission lines.
In the following, we will use country and bidding zone interchangeably. Power supply is dominated by variable renewables (solar PV, wind onshore and offshore, run of river) and complemented by carbon-free thermal power plants (biomass, H$_2$, nuclear and waste) and different storage technologies (batteries, H$_2$ caverns, hydro reservoirs with natural inflow and pumped hydro storage). Power demand represents 2050 projections of conventional electricity demand as well as load from flexible electric vehicle charging, heat pumps and electrolysers. The latter are built to supply 2050 industrial H$_2$ demand projections, which can also be covered by imports from outside the EU for a constant price. Detailed data and technology assumptions are described in \cite{johanndeiter2025b}. Importantly, we consider three wind onshore profiles in each bidding zone that differ in terms of their capacity factor time series. Existing wind onshore power plants are represented by the country average. For wind onshore investment options in each bidding zone, a profile named \textit{High FLH} incurs higher full load hours (FLH), but a lower market value (MV) than the second profile \textit{High MV}, such that the latter can  be considered more system-friendly (cf. \cite{johanndeiter2025} for details). 

We employ a sequential optimisation approach starting with an investment optimisation of four scenarios, each resulting in a different capacity mix. All scenarios share the assumption that the fully decarbonised European energy system will be dominated by variable renewables and build on 2030 national capacity expansion targets. Technically, these assumptions imply that all scenarios start from a brownfield energy system with existing carbon-free electricity generation capacities that represent 2030 national targets. Additional electricity and H$_2$ generation capacities are determined by minimising system costs under the consideration of a European target share of variable renewables (solar, wind and run of river) of annual electricity demand (including storage losses) of at least 95\%, enforced by a respective constraint. The optimisation is conducted based on sample weeks of hourly capacity factor and load time series from the weather year 2019, which is considered average. The four scenarios vary in terms of two variables that affect the (i) level and (ii) technological composition of capacities. Particularly, we vary (i) the H$_2$ import price that drives the level of capacities installed in the EU; and (ii) cost assumptions for solar PV and battery that influence the distribution of capacities between solar PV and wind power (cf. Table \ref{tab:tab_scen_i}).

\begin{figure}[!ht]
\centerline{\includegraphics[width=0.9\columnwidth]{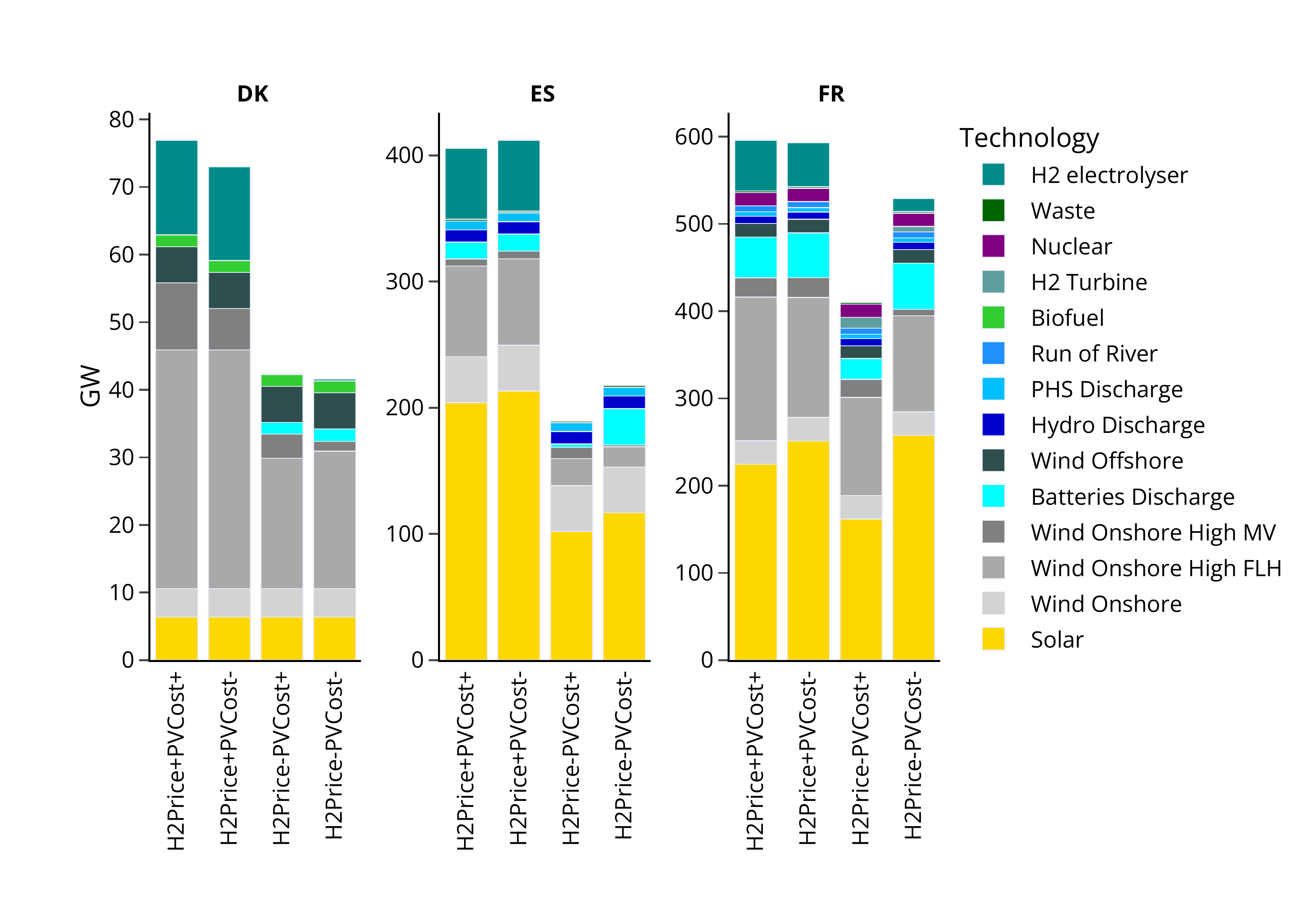}}
\caption{Electricity and hydrogen generation capacities installed in four investment scenarios in Denmark (DK), Spain (ES) and France (FR).}
\label{fig:capas}
\end{figure}

For our analysis we focus on three bidding zones with different characteristics: Denmark (DK), Spain (ES) and France (FR). Figure \ref{fig:capas} shows capacities resulting from the four investment optimisations: In terms of variable renewable capacities, Spain is dominated by solar PV, Denmark by wind onshore and France's capacity mix is rather balanced between the two technologies. In Spain, variable renewable generation is complemented by hydro storage, in Denmark by biofuel power plants and in France by nuclear generators and hydrogen turbines.

\begin{table}[h!]
\caption{Overview of varying assumptions for investment scenarios}
\footnotesize
\label{tab:tab_scen_i}
\centering
\begin{threeparttable}
\begin{tabular}{c c c c}
\toprule
& & \multicolumn{2}{c}{H\textsubscript{2} Price (\euro{}/MWh)} \\
\cmidrule(lr){3-4} \makecell{PV cost\\(\euro{}/GW)} & \makecell{Battery cost\\(\euro{}/GW)} & 45.07 & 116.9 \\
\midrule
308 & 220 & H2Price--PVcost-- & H2Price+PVcost-- \\
400 & 290 & H2Price--PVcost+ & H2Price+PVcost+ \\
\bottomrule
\end{tabular}
\begin{tablenotes}[flushleft]
\scriptsize
\item Sources: \citet{ENS2020}, \citet{Tyndp2022}, \citet{brandle2021}.
\end{tablenotes}
\end{threeparttable}
\end{table}

Hourly dispatch of the four resulting capacity mixes is optimised for nine full-year scenarios each. The nine scenarios vary in terms of variables that particularly affect market prices: three alternative H$_2$ import price assumptions\footnote{Initially we used all three H$_2$ price assumptions for the investment scenarios, but found no significant difference in capacities between the H2Price+ and H2Price++ scenario.} in combination with the weather years 2010, 2015 and 2019 (cf.\ Tables \ref{tab:tab_scen_d} and \ref{tab:tab_scen_d2}). 

Weather year 2010 is characterised by relatively low renewable capacity factors and high peak demand, whereas 2015 represents the opposite. To mimic behavior of myopic market actors, we apply a rolling-horizon approach that sequentially optimises each day in hourly resolution with the remaining days aggregated at an increasingly coarse resolution in the forward-looking window. In total, our approach results in 36 scenario outcomes.

\begin{table}[h!]
\caption{Overview of varying assumptions regarding weather years for the dispatch scenarios. Capacity factors are averaged across Europe.}
\footnotesize
\centering
\label{tab:tab_scen_d}
\begin{threeparttable}
\begin{tabular}{c c c c c c c c}
\toprule
 \multirow{2}{*}{\makecell{Weather\\year}} & \multicolumn{5}{c}{Average capacity factor (\%)} & \multicolumn{2}{c}{Electric load} \\
\cmidrule(lr){2-6} \cmidrule(lr){7-8} & Offshore  & Onshore & \makecell{Onshore\\High FLH} & \makecell{Onshore\\High MV} & PV &  Total (TWh) & Peak (GW) \\
 \midrule
2010	&	35.2	&	23.2	&	30.9	&	26.0	&	12.7	&	4286	&	695	\\
2015	&	39.1	&	26.1	&	34.7	&	29.6	&	13.0	&	4107	&	644	\\
2019	&	34.3	&	25.3	&	34.4	&	28.8	&	13.0	&	4110	&	619 \\
\bottomrule
\end{tabular}
\begin{tablenotes}[flushleft]
\scriptsize
\item Sources: \citet{staffel2016}, \citet{Tyndp2022}.
\end{tablenotes}
\end{threeparttable}
\end{table}

\begin{table}[h!]
\caption{Overview of varying assumptions regarding H\textsubscript{2} prices for the dispatch scenarios.}
\centering
\label{tab:tab_scen_d2}
\footnotesize
\begin{threeparttable}
\begin{tabular}{l c}
\toprule
\makecell{H$_2$ price\\level} & \makecell{H$_2$ price\\(\euro{}/MWh)}  \\
\midrule
H2Price-- & 45.07 \\
H2Price+ & 116.90 \\
H2Price++ & 188.73 \\
\bottomrule
\end{tabular}
\begin{tablenotes}[flushleft]
\scriptsize
\item Sources: \citet{Tyndp2022}, \citet{brandle2021}, own assumption.
\end{tablenotes}
\end{threeparttable}
\end{table}

\begin{figure}[!ht]
\centerline{\includegraphics[width=0.8\columnwidth]{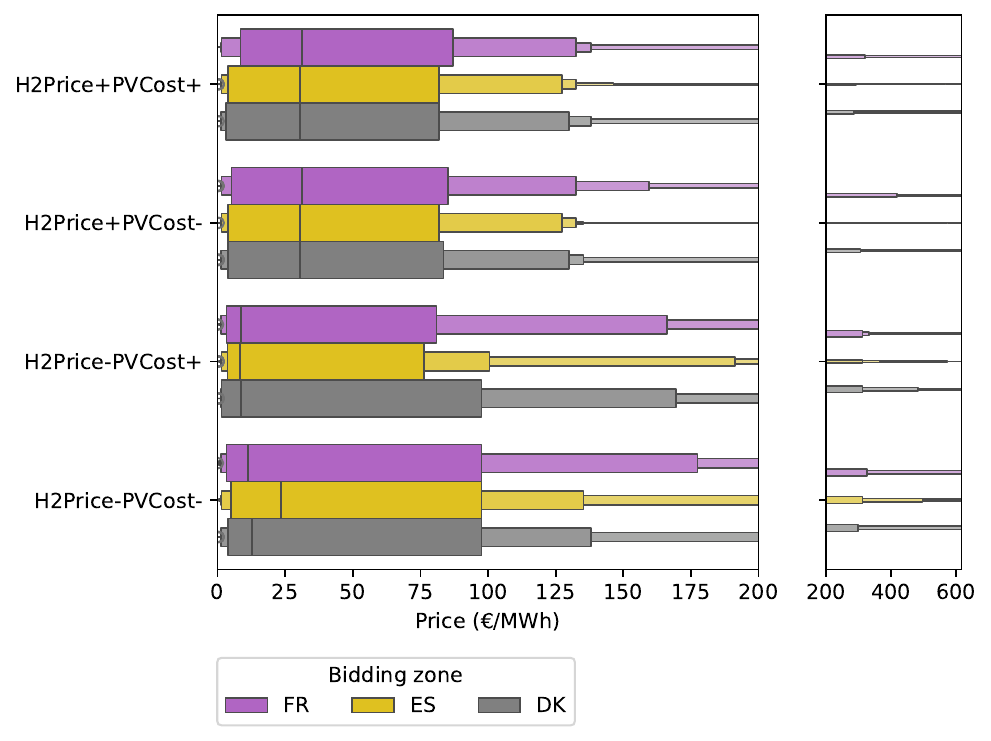}}
\caption{Electricity price distribution for each investment scenario each across 9 dispatch scenarios in Denmark (DK), Spain (ES) and France (FR) with prices capped at 617\,\euro/MWh. Lines represent the median, the largest box spans from the 25\% and 75\%-quartiles and successively smaller boxes from 12.5\%– to 87.5\%-percentiles, from 6.25\%- to 93.75\%-percentiles and so on.}
\label{fig:prices}
\end{figure}

Figure \ref{fig:prices} shows a letter-value-plot \citep{hofmann2017} of electricity price distributions resulting from the nine dispatch scenarios for each invest scenario.
Overall, prices differ more strongly between investment scenarios than between countries. Across all scenarios, more than 75\% of all prices are below 100\,\euro/MWh. Differences are mainly driven by the H\textsubscript{2} import price assumption of the investment scenario: If it is high, median prices are approximately 35\,\euro/MWh; when it is low, median prices range between 10 and 25\,\euro/MWh. 

Due to our sequential optimisation approach, we observe involuntary load shedding at a maximum price of 4000\,\euro{}/MWh across all scenarios, particularly those that represent the weather year 2010. However, the energy crisis of 2022 revealed that such extremely high electricity prices are politically undesired, when revenues were capped across Europe \citep{eucom2022}. Therefore, we cap electricity prices at the value of lost load of the most expensive industrial demand response unit in our model, but also conduct a sensitivity analysis without the cap (see \ref{app}).

\subsection{Optimal strike prices under uncertainty}
\label{sec:method_sp}
\begin{figure}[!ht]
\begin{subfigure}{0.48\linewidth}
    \includegraphics[width=\columnwidth]{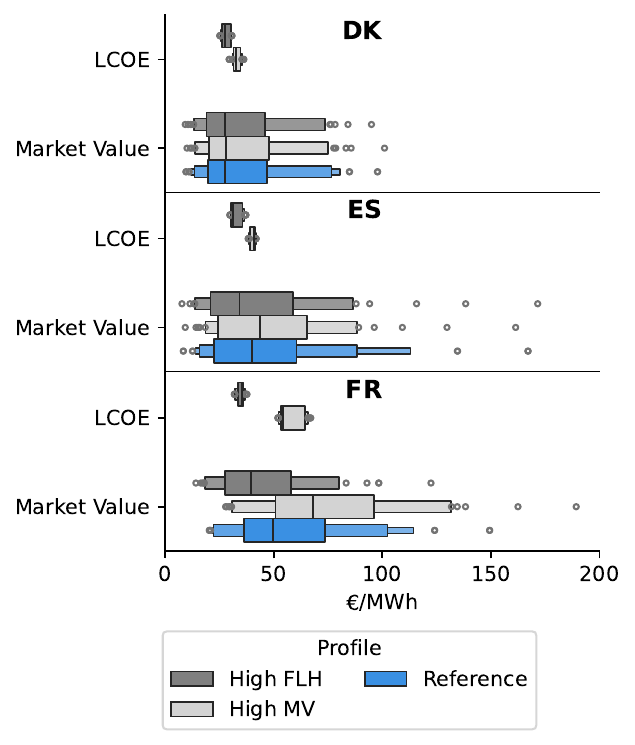}
    \caption{Production-based strike price components $\text{lcoe}_{i,n,s}$ and $v_{i,n,s}$ (High FLH, High MV) as well as $v_{n,s}$ (Reference).}
\end{subfigure}
\hfill
\begin{subfigure}{0.48\linewidth}
    \includegraphics[width=\columnwidth]{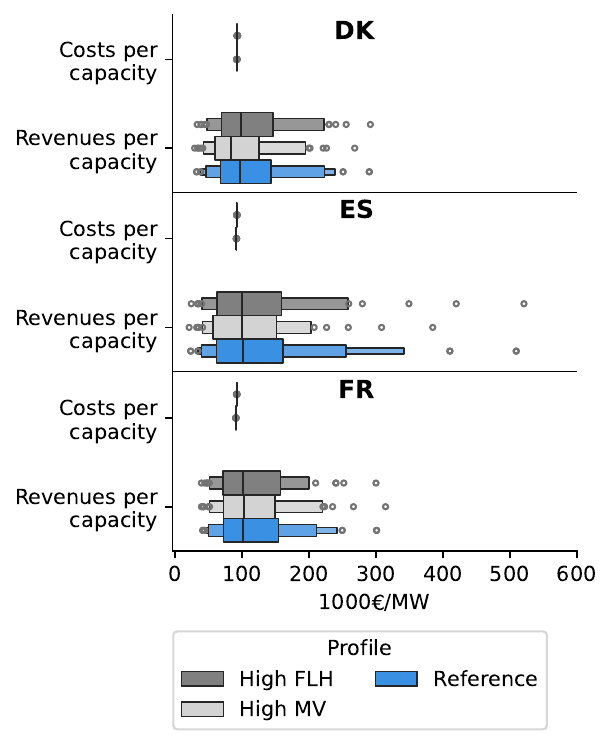}
    \caption{Capacity-based strike price components $C_{i,n,s}/Q_{i,n,s}$ and $r_{i,n,s}$ (High FLH, High MV) as well as $r_{n,s}$ (Reference).}
\end{subfigure}
\caption{Distribution of production-based and capacity-based strike price components across 36 dispatch scenarios for two power plants and the reference power plant in Denmark (DK), Spain (ES) and France (FR). Lines represent the median, the largest box spans from the 25\% and 75\%-quartiles and successively smaller boxes from 12.5\%– to 87.5\%-percentiles, and so on.}
\label{fig:spc_both} 
\end{figure}

We assume the optimised dispatch results to represent investors' expectations on possible market outcomes at the time of the auction for the Contracts for Difference. Hence, we use the results of all 36 dispatch scenarios $s \in S$ to determine optimal strike prices under uncertainty $\tilde{S}_{i,n}^\text{type}$ for each CfD type and both wind power plants $i \in I=\{\text{High FLH}, \text{High MV}\}$ in each bidding zone $n \in N$ based on Equations (\ref{eq:sbasic_unc}), (\ref{eq:s2way_unc}) and (\ref{eq:sfin_unc}).
The distribution of the components used to determine optimal strike prices are depicted in Figure \ref{fig:spc_both}.
The cost-based components, LCOE and costs per capacity, are only affected by volume risk, and thus have lower volatility than the revenue-based components, market value and revenues per capacity, which also reflect price risk.
As variable costs of wind power plants are low, the cost per capacity even has very little volume risk.
The volatility of revenue-based components is largely driven by weather years. The year 2010 in particular has significantly higher market values and revenues than the other two years and also higher volatility. Therefore, we study a sensitivity analysis without the year 2010.
The difference in FLH, thus in LCOE, between the High FLH and High MV profiles is highest in France, followed by Spain and then Denmark.
The difference in the production-based market values follows this trend, but the difference in revenues per capacity does not -- the latter is highest in Denmark. 

As a result, we consider all 36 scenarios to form one optimal strike price for each wind power plant in all bidding zones and for all three CfD types. 

\subsection{Ex post analysis of CfD payments, cost recovery and consumer cost}
Based on the optimal strike prices, we examine ex post CfD payments, wind power plants' cost recovery and consumer prices in each scenario realisation.
First, we calculate each wind onshore power plant's cost recovery $\Psi_{i,n,s}^\text{type}$ as the ratio of market revenues and CfD payments to annual costs resulting ex post from each scenario:
\begin{align}
\label{eq:ex_post_profit}
  \Psi_{i,n,s}^\text{type}=\frac{\sum_{t=1}^{T}(q_{t,i,n,s}p_{t,n,s})+P^\text{type}_{i,n,s}}{ C_{i,n,s}} . 
\end{align}

For instance, for the basic CfD we calculate
\begin{align}
\label{eq:profit_basic}
 \Psi_{i,n,s}^\text{basic}=\frac{\sum_{t\in T}(q_{t,i,n,s}p_{t,n,s})+\sum_{t=1}^{T}q_{t,i,n,s}\,(\tilde{S}_{i,n}^\text{basic}-p_{t,n,s})}{C_{i,n,s}}.
\end{align}

As a measure of volatility, we determine the coefficient of variation \citep[~p. 73]{sachs2006} of cost recovery for each power plant, bidding zone and CfD type as 
\begin{align}
\label{eq:rev_vol}
 \sqrt{{\frac{1}{S}\sum_{s\in S}\left(\Psi^\text{type}_{i,n,s} - \overline{\Psi}^\text{type}_{i,n}\right)^2}} \bigg/ \overline{\Psi}^\text{type}_{i,n}\,,
\end{align}
where $\overline{\Psi}^\text{type}_{i,n}=\frac{1}{S}\sum_{s\in S}\Psi_{i,n,s}^\text{type}$ is the mean of the cost recovery.

Second, we assess consequences for consumers by analysing average consumer price $\Phi_{n,s}^\text{type}$ in bidding zone $n$ as the sum of the volume-weighted electricity price and a uniform levy that finances CfD payments,
\begin{align}
\Phi^\text{type}_{n,s}=\frac{\sum_{t\in T}p_{t,n,s} \cdot d_{t,n,s}}{\sum_{t\in T}d_{t,n,s}} + \frac{P^\text{type}_{n,s}}{\sum_{t\in T}d_{t,n,s}},
\end{align}
where $d_{t,n,s}$ is the hourly electricity demand in bidding zone $n$ and scenario $s$.
We also calculate the coefficient of variation of the total consumer prices
\begin{align}
\label{eq:cost_vol}
 \sqrt {\frac{1}{S}\sum_{s\in S}\left(\Phi^\text{type}_{n,s} - \overline{\Phi}^\text{type}_{i,n}\right)^2} \bigg/\overline{\Phi}^\text{type}_{i,n},
\end{align}
where $\overline{\Phi}^\text{type}_{i,n}=\frac{1}{S}\sum_{s\in S}\Phi_{i,n,s}^\text{type}$.
\newpage
\section{Results and discussion}
\label{sec:results}
We first present numerical results for optimal strike prices under uncertainty for two wind power plants in three bidding zones by CfD type. Then, we analyse the distribution of the power plants' cost recovery and consumer prices resulting ex post across all 36 scenarios.

\subsection{Optimal strike prices}

\begin{figure}[h!]
\centerline{\includegraphics[width=0.9\columnwidth]{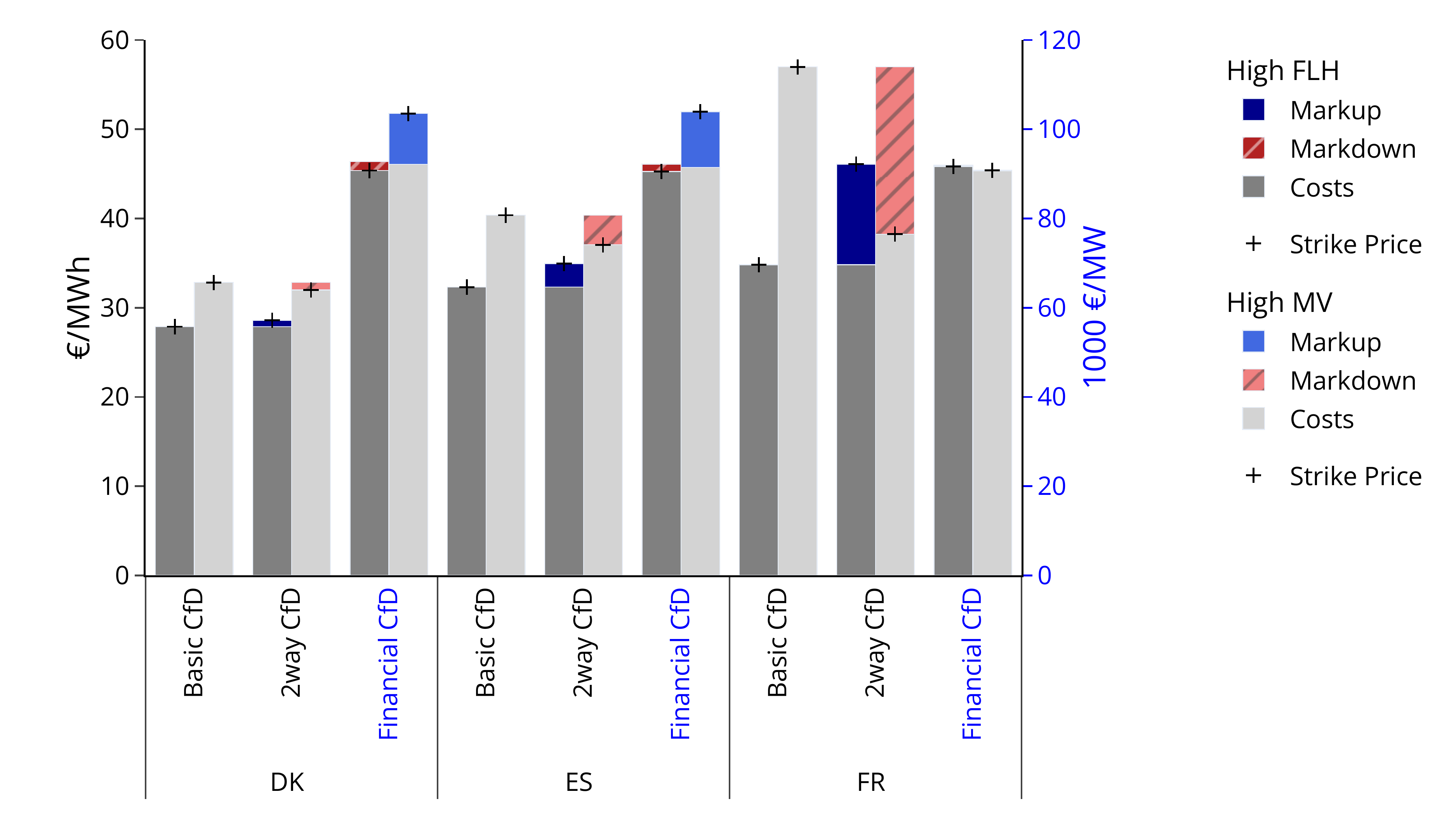}}
\caption{Optimal strike prices under uncertainty for two wind onshore power plants (High FLH, left bar and High MV, right bar) in Denmark (DK), Spain (ES) and France (FR). Differences between the strike price and expected costs (expected value of LCOE for basic and 2way CfD, expected value of total cost per capacity for financial CfD, according to Equations (\ref{eq:sbasic_unc}) to (\ref{eq:sfin_unc})) is called markup if positive and markdown if negative. Note that strike prices for the financial CfD are shown on the right y-axis in units of 1000\,\euro/MW.}
\label{fig:sp}
\end{figure}

Figure \ref{fig:sp} depicts optimal strike prices under uncertainty by wind power plant -- as distinguished by generation profile --, bidding zone and CfD type.
In line with the theory in Section \ref{sec:theory}, strike prices of the basic CfD equal expected LCOE.
As these are lower for the High FLH profile by design in every country, this CfD design does not incentivise system-friendly design and siting.
For the 2way and financial CfDs, the differences between strike prices and expected costs (per generation or per capacity) are indicated as markups (if positive) or markdowns (if negative).
The 2way CfD results in markups for the High FLH and markdowns for the High MV profile in all three countries. Markups and markdowns reflect the difference in a power plant's individual market value and the bidding zone's average, i.e.\ the reference price: an above-average market performance allows a power plant to decrease the strike price below LCOE by a markdown and requires an increase by a markup in the opposite case.
Therefore, the 2way CfD consistently incentivises the more system-friendly High MV profile with the effect size driven by the difference in market values. 
The difference between profiles, both in terms of LCOE and market value, is lowest in Denmark and highest in France. Here, the markup and markdown are so high that the optimal strike price of the High FLH profile is above the optimal strike price of the High MV profile.

By design, the expected cost per capacity, which serves as baseline for the optimal strike price of the financial CfD, is nearly equal for all countries and scenarios and only differs in terms of total variable costs. Hence, strike prices only differ in terms of their markups or markdowns. We observe markups for High MV and markdowns for High FLH profiles, illustrating that the financial CfD does not offer the same siting and design incentive in favour of the High MV profile as the 2way CfD, because a higher market value, i.e.\ higher revenues per generation, does not necessarily imply higher revenues per capacity (cf.\ Figure \ref{fig:spc_both}). 

\subsection{Ex post cost recovery}
\begin{figure}[!ht]
\centerline{\includegraphics[width=1\columnwidth]{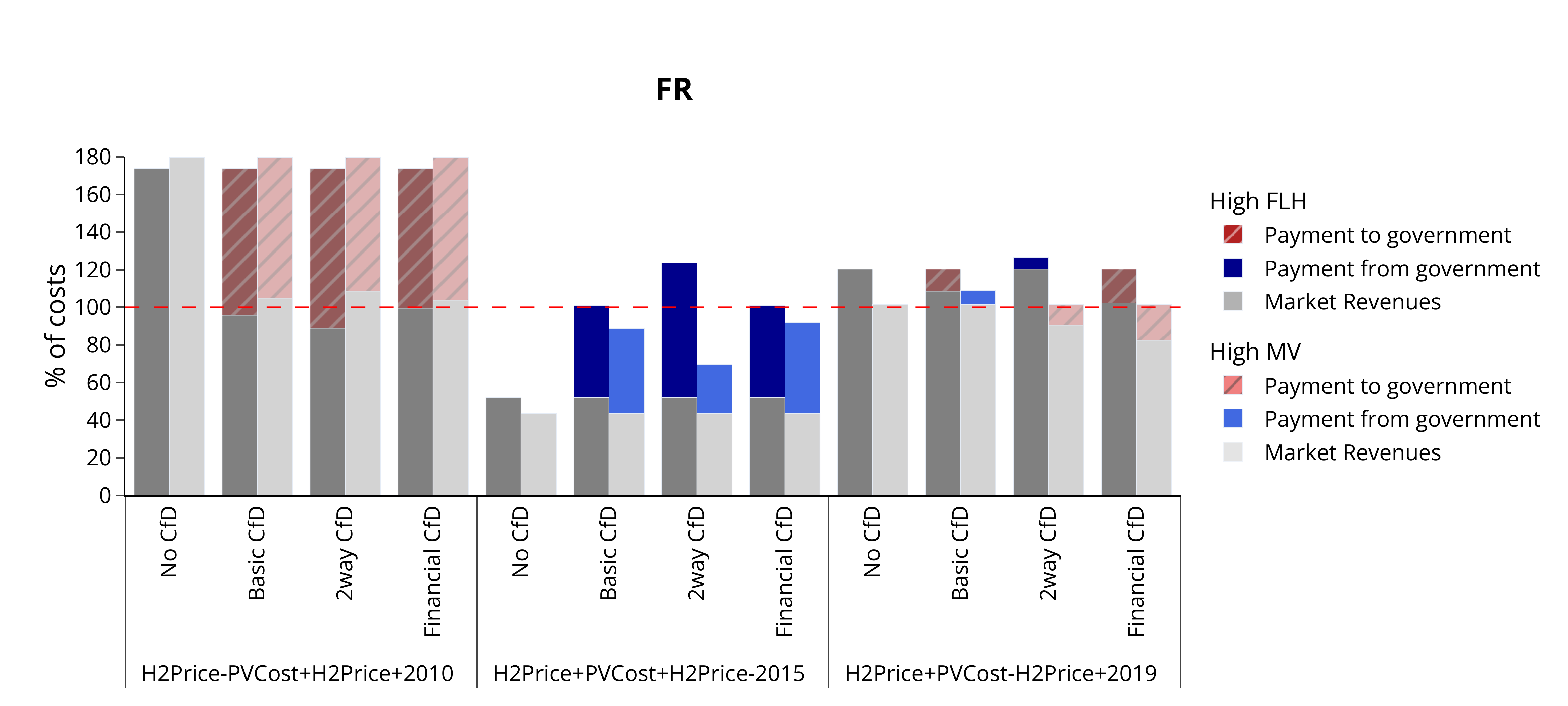}}
\caption{Ex post cost recovery across three selected scenarios in France. Scenario names are composed of the two determinants of the investment scenario variation followed by the two determinants of the dispatch scenario variation.}
\label{fig:ex_post_FR}
\end{figure}

Figure \ref{fig:ex_post_FR} shows ex post cost recovery for three selected scenarios in France.
The scenarios illustrate three characteristic cases: (i) high cost recovery from market revenues without the CfD, where both wind power plants make significant payments \emph{to} the government for any CfD,
(ii) low cost recovery from market revenues, where all power plants receive varying amounts of payments from the government for all CfD types, and
(iii) cost recovery from market revenues around 100\% without a CfD, where payments from or to the government are comparably low and made depending on CfD type and power plant.

\begin{figure}[!h]
\centerline{\includegraphics[width=0.55\columnwidth]{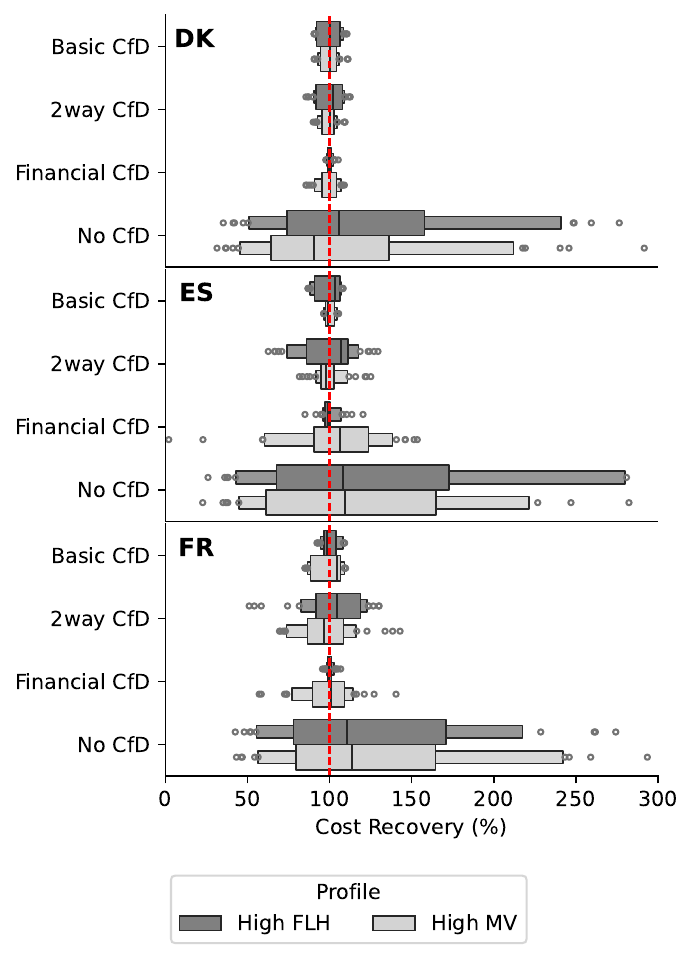}}
\caption{Distribution of ex post cost recovery across 36 dispatch scenarios for two wind power plants in Denmark (DK), Spain (ES) and France (FR). Plot is truncated at 300\% excluding 2 observations in FR and 5 in ES.}
\label{fig:ex_post}
\end{figure}

\begin{figure}[!h]
\begin{subfigure}{0.48\linewidth}
    \centerline{\includegraphics[height=0.7\columnwidth]{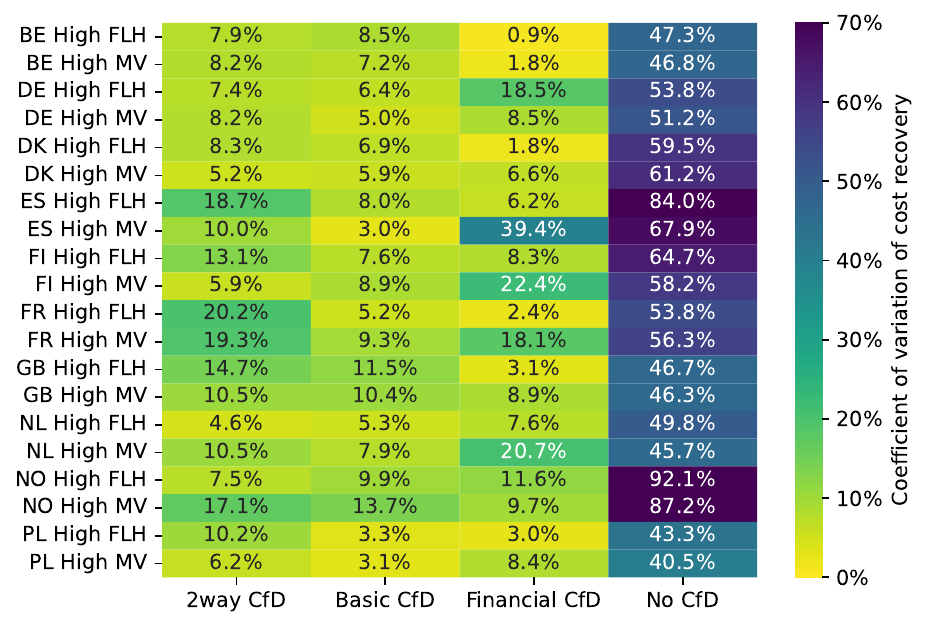}}
    \caption{Coefficient of variation (CV) of cost recovery by CfD type for 20 wind power plants located in 10 bidding zones specified by two-digit country codes. The color code is truncated at 70\%}
     \label{fig:heatmap_costrec}
\end{subfigure}
\begin{subfigure}{0.48\linewidth}
    \centerline{\includegraphics[height=0.7\columnwidth]{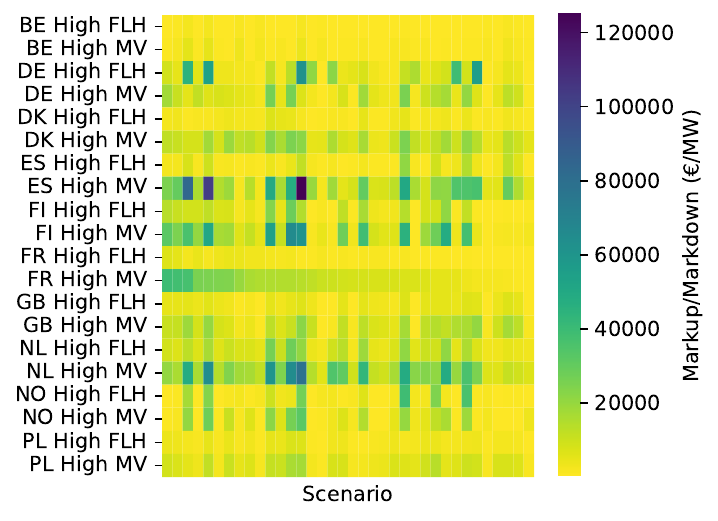}}
    \caption{Difference in revenues per capacity  between individual and reference profile (markup/markdown) for 20 wind power plants located in 10 bidding zones specified by two-digit country codes.}
    \label{fig:rev_diff}
\end{subfigure}
\caption{Cost recovery variation and revenue differences by power plant.}
\label{fig:heat_maps_recovery}
\end{figure}

More comprehensively, Figure \ref{fig:ex_post} shows the distribution of ex post recovery for all three bidding zones across all 36 scenarios.
The following key observations can be made.
First, any type of CfD decreases the spread of the cost recovery distribution significantly compared to a situation, where power plants solely rely on market revenues (``no CfD'').
This holds for the illustrated 25\%...75\% and 12.5\%...87.5\% percentiles as well as outliers across all countries.
It is also confirmed by the coefficients of variation (CVs) shown in Figure \ref{fig:heatmap_costrec} for all bidding zones with endogenous investments in both profiles.

Second, when comparing the different CfD types, the basic CfD mostly has a lower CV than the 2way CfD.
Moreover, Figure \ref{fig:ex_post} shows that for the High FLH profile, the financial CfD achieves the lowest spread of cost recovery for all three countries considered.
However, this neither holds for the High MV profile, nor for every country individually.
And indeed the CV is lowest under the financial CfD for several power plants, 
but highest for the others.
This can be explained by the difference in revenues per capacity between the individual power plant and the reference shown in Figure \ref{fig:rev_diff} for 20 power plants and all scenarios.
If the difference is high in some scenarios, like for the High MV profile in ES, Finland (FI) and the Netherlands (NL), then the variation of cost recovery increases significantly driven by the revenue side.
If the difference in revenues per capacity is low, however, as is the case for both profiles in Belgium (BE), then costs are the main driver of cost recovery. And as shown in Figure \ref{fig:spc_both} in Section \ref{sec:method_sp}, costs vary less when divided by capacity than when divided by generation (as for the basic and 2way CfDs).

Third, while reducing the volatility of cost recovery, the CfDs do not generally increase the cost recovery overall.
With the exception of the High MV profile in Denmark, the median cost recovery is above 100\% without any CfD and decreases with a CfD in a majority of cases. This result highlights that the purpose of CfDs in highly renewable electricity markets is to reduce revenue volatility rather than increasing average revenues.

\subsection{Consumer prices}
\begin{figure}[!ht]
\centerline{\includegraphics[width=0.5\columnwidth]{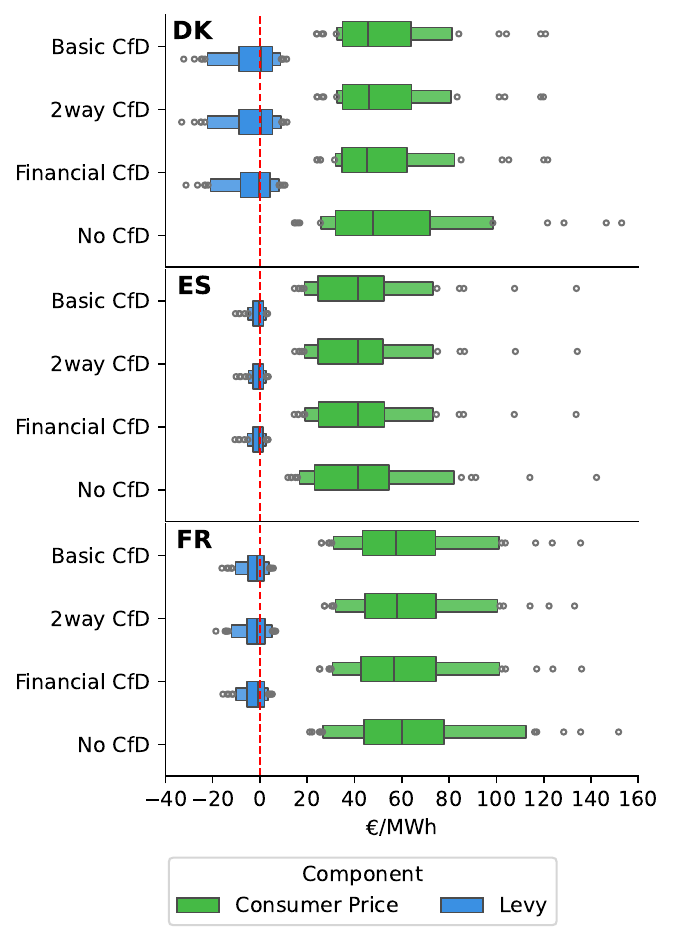}}
\caption{Distribution of consumer price composed of electricity price (not depicted) and levy across 36 dispatch scenarios in Denmark (DK), Spain (ES) and France (FR).}
\label{fig:consumercosts}
\end{figure}

\begin{figure}[!ht]
\centerline{\includegraphics[width=0.5\columnwidth]{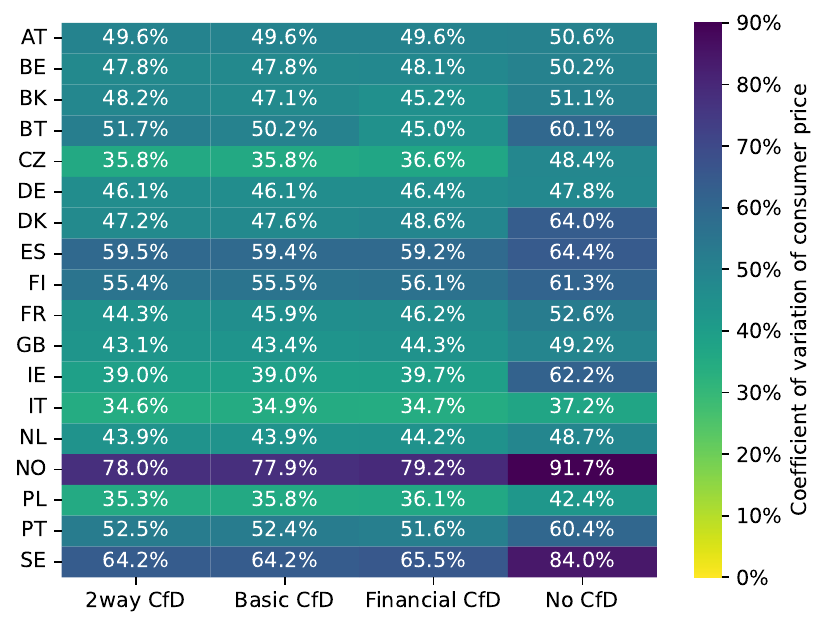}}
\caption{Coefficient of variation of total consumer price by CfD type for all 19 bidding zones specified by two-digit country codes.}
\label{fig:heatmap_cons}
\end{figure}

The distribution of the total consumer price and total levies per unit of electricity in the three bidding zones DK, ES and FR is shown in Figure \ref{fig:consumercosts}.
Figure \ref{fig:heatmap_cons} shows the corresponding coefficients of variation of the consumer price for all bidding zones. 
Overall, there are significant variations of average consumer price and levies between countries.
However, there are no significant differences between CfD types and only a small decrease of the consumer price spread through introduction of any CfD as compared to no CfD.
In line with the observation that all three CfDs do not generally increase the overall cost recovery, the median levy is close to zero in all cases.
The highest coefficients of variation in Norway, Spain and Finland can be explained by a higher range of consumer prices due to several scenarios with very low prices.

\subsection{Sensitivity analysis}
To test the robustness of these results, we conduct a sensitivity analysis.
We determine both the cost recovery and consumer prices for two additional cases, (i) without the cap on extreme prices and (ii) with the weather year 2010 removed from the set of dispatch scenarios.
The results are shown in \ref{app}.
Without the price cap, the volatility and overall level of cost recovery is increased, in particular in Denmark and France. The same holds for the volatility and level of consumer prices.
Without the year 2010, the volatility and overall level of both cost recovery and consumer prices are decreased.
However, in both cases, our findings on the effect of CfDs in general and the comparison of the three CfD types remain unchanged.

\subsection{Limitations}
Our analysis is subject to two main limitations. First, it is important to highlight that our results only concern the volatility of cost recovery and consumer prices, and not their level, which would be driven by several factors that we neglect in our analysis. To begin with, we assign an equal weight to each scenario considered for the calculation of expected values. Hence, our derived strike prices rely on the assumption that investors are risk-neutral. Strike prices by risk-averse investors, in turn, would typically reflect the likelihood and volatility of cost recovery by CfD type either by a risk premium \citep{neuhoffetal2022} or by assigning a higher weight to market scenarios with low profits. In this case, strike prices and therefore, consumer prices, for the 2way CfD are likely to be higher than for the basic CfDs, if investors are risk-averse. Moreover, auctions for renewable support schemes have historically not always been competitive. Germany, for instance, experienced undersubscribed auctions for wind onshore support resulting in average strike price close to the auction's price ceiling \citep{bnetza2023}. Conversely, offshore wind auctions were highly competitive, with several investors submitting bids requiring no financial support \citep{bnetza2023b}. Lastly, we do not take into account the reduction in capital costs induced by the reduction of revenue volatility by all CfD types as documented by previous work (cf. Section \ref{sec:intro_design}). This reduction would affect investments and lower system costs \citep{bachneretal2019,helistoe2019}, which in turn would be reflected in lower prices for consumers. Similarly, we neglect the effect of CfD design on investment and operational decisions as discussed in Section \ref{sec:intro_design}. A complete evaluation of CfD design should take into account a combined assessment of these effects. 

Second, we only consider a limited set of scenarios, consequently, a limited set of risks. Particularly, we focus on a subset of weather and technological risk, represented by three weather years, varying hydrogen import prices and investment costs of competing variable renewables. \citet{jimenez2024} and \citet{johanndeiter2025b} show that weather variations and the hydrogen import price are indeed the main drivers of price and volume risk in sector-coupled, highly renewable electricity markets. Yet, previous work indicates that large-scale energy system models tend to structurally underestimate price volatility in the day-ahead market and thus, the underlying price risk that could affect total revenue uncertainty \citep{terhorstetal2026}. 
Further technological uncertainties could arise from climate change, the degree of electrification in different sectors, total demand as well as its level of flexibility, or the invention of disruptive technologies. Unlike weather and short-term price fluctuations, the extent of these uncertainties is itself uncertain with potentially significant effects on market outcomes. Additionally, our analysis neglects regulatory uncertainty. Our sensitivity analysis, where we remove the price cap from market outcomes, reveals that our results on the impact of CfD design are robust to the introduction of price caps. In a more realistic market setting, such price caps could result from capacity remuneration mechanisms that would additionally affect electricity generation capacities. Therefore, the interaction of CfDs and capacity remuneration mechanisms should be subject to further research. Finally, despite their relevance for cost of capital \citep{leveque2026}, we believe that more disruptive regulatory risks, such as a decline in climate policy ambitions or international cooperation, must not be reflected in a renewable support scheme and are therefore out of scope of this study. 

\section{Conclusion}
\label{sec:conclusion}
Two-way CfDs can mitigate the price and volume risks of electricity consumers and investors.
In this paper, we studied and compared the risk mitigation effect of three CfD designs, the production-based basic and 2way CfDs as well as the capacity-based financial CfD.
We started by theoretically deriving optimal strike price bids under uncertainty for three different CfD types based on investors' expectations. 
Then, we assumed a set of highly renewable European electricity market scenario results from a sector-coupled power system model to represent those expectations and calculated resulting strike prices and CfD payments.
Finally, we analysed the distribution of investors' cost recovery for wind turbines and consumer prices, both including CfD payments, across scenarios and countries.

We find that, first, all three CfD types significantly reduce the volatility of the profit-to-cost-ratio (cost recovery) compared to full market price exposure.
Second, consumer price volatility is also reduced by all CfD types, but to a much lesser extent.
Third, we find no significant difference in consumer price volatility between the three CfD types considered.
Fourth, the lowest cost recovery volatility is achieved by the basic CfD and the financial CfD with a reference power plant similar to the investor's one.
In contrast, the CfD with highest cost recovery volatility is the financial CfD with a large difference between the reference price and the investor's own market revenues.
Existing literature shows that this CfD design does not cause any dispatch distortions and, to a certain extent, incentivises system-friendly investments.
Overall, our results confirm previous work that designing CfDs is subject to a trade-off between incentivising system-friendliness and reducing risk within a novel framework to assess CfD design under uncertainty.

In conclusion (cf.\ Table \ref{tab:conclusion}), our work suggests that the financial CfD is the most suitable one to support the energy transition in highly renewable electricity systems.
On the one hand, as already found by other authors, the financial CfD is the only one among the three CfD types incentivising both optimal dispatch and, to a certain extent, also system-friendly siting.
On the other hand, our novel analysis of different CfD types under uncertainty shows that, together with the basic CfD, it also offers the lowest cost recovery volatility for investors, if the reference plant is similar to the investor's one.
In this case, however, it offers lower incentives for system-friendly siting.
Regarding consumer price volatility, it performs equally well as the other CfD types considered. 

\begin{table}[]
    \centering
    \footnotesize
    \caption{Comparison of CfD types. The first two rows regarding system-friendly siting and optimal dispatch are known from the literature, while the last two rows are new insights generated in this paper.}
    \begin{tabular}{l c c c c}
        \toprule
         & No CfD & Basic CfD & 2way CfD & Financial CfD \\
         \midrule
         Optimal dispatch & + & -- & -- & + \\
         System-friendly investment & + & -- & + & $\circ$ \\
         \midrule
         Profit volatility & -- & + & $\circ$ & + / $\circ$ \\
         Consumer price volatility & -- & $\circ$ & $\circ$ & $\circ$ \\
         \bottomrule
    \end{tabular}
    \label{tab:conclusion}
\end{table}

Future work should address the choice of reference profiles for financial CfDs, which strongly drives profit volatility. 
For instance, investors may be allowed to choose the reference plant according to their individual risk preferences.
Risk-averse investors could then use a very similar power plant, while risk-loving investors could choose the average of their respective bidding zone.
A CfD like this would require a thorough design, e.g.\ to avoid potential net payments from the government to investors.
Moreover, future work should study interactions with other regulatory instruments, such as capacity remuneration mechanisms. \\

\textbf{CRediT author statement} \\
\hspace*{6pt} \textbf{Silke Johanndeiter}: Conceptualization of this study, Methodology, Data curation, Writing - Original Draft, Visualization, Formal analysis. \textbf{Jonas Finke}: Methodology, Writing - Original Draft. \textbf{Justus Heuer}: Writing - Original Draft.  \\ 

\textbf{Acknowledgements} \\
\hspace*{6pt} We would like to thank Masooma Fazili for supporting us in data curation. \\ \\
\hspace*{12pt} \textbf{Declaration of AI-assisted technologies in the writing process}   \\ 
\hspace*{6pt} During the preparation of this work the authors used ChatGPT in order to support data visualisation code and language proofing. After using this tool, the authors reviewed and edited the content as needed and take full responsibility for the content of the publication. \\ \\
\hspace*{12pt} \textbf{Data availability statement} \\
\hspace*{6pt} The data and code that support the findings of this study are openly available in public repositories at \href{https://github.com/TradeRES/TradeRES-Backbone-demo}{https://github.com/TradeRES/TradeRES-Backbone-demo}, \href{https://doi.org/10.5281/zenodo.10829706}{https://doi.org/10.5281/zenodo.10829706} and \href{https://doi.org/10.5281/zenodo.17986036}{https://doi.org/10.5281/zenodo.17986036}. \\ 

\newpage
\FloatBarrier
\begin{singlespace}
\bibliographystyle{elsarticle-harv} 
\bibliography{references.bib}
\end{singlespace}

\newpage
\FloatBarrier
\appendix

\section{Supplementary Results}
\label{app}
\setcounter{figure}{0}

\begin{figure}[!ht]
\begin{subfigure}[t]{0.48\linewidth}
    \includegraphics[width=\columnwidth]{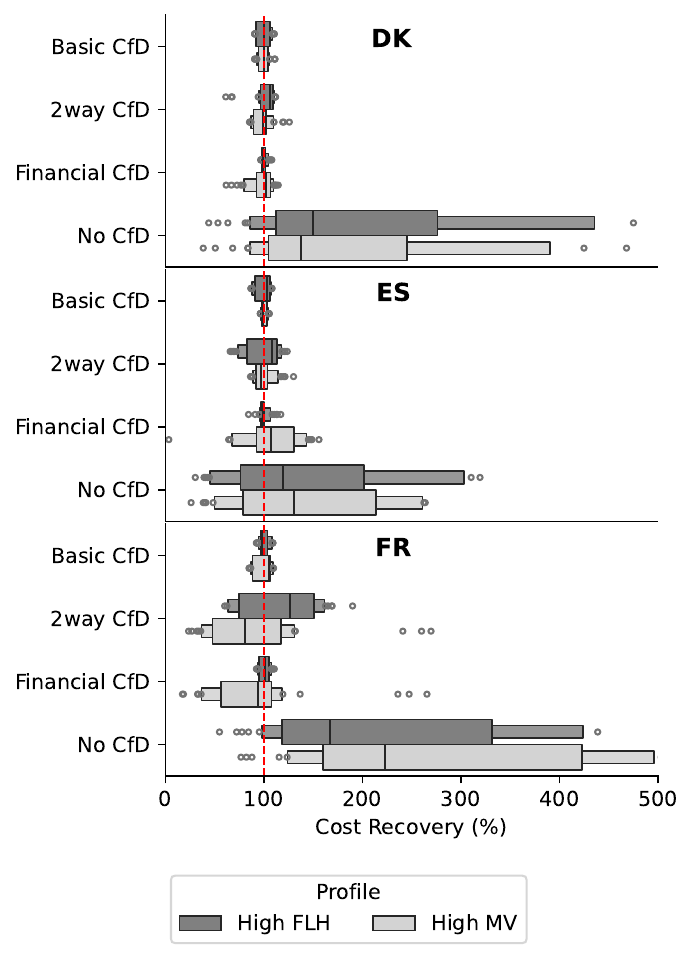}
    \label{fig:ex_post_uncapped}
    \caption{Distribution of ex post cost recovery across 36 dispatch scenarios without a price cap for two wind power plants in Denmark (DK), Spain (ES) and France (FR). Plot is truncated at 500\% excluding 6 observations in each bidding zone.}
\end{subfigure}
\hfill
\begin{subfigure}[t]{0.47\linewidth}
    \includegraphics[width=\columnwidth]{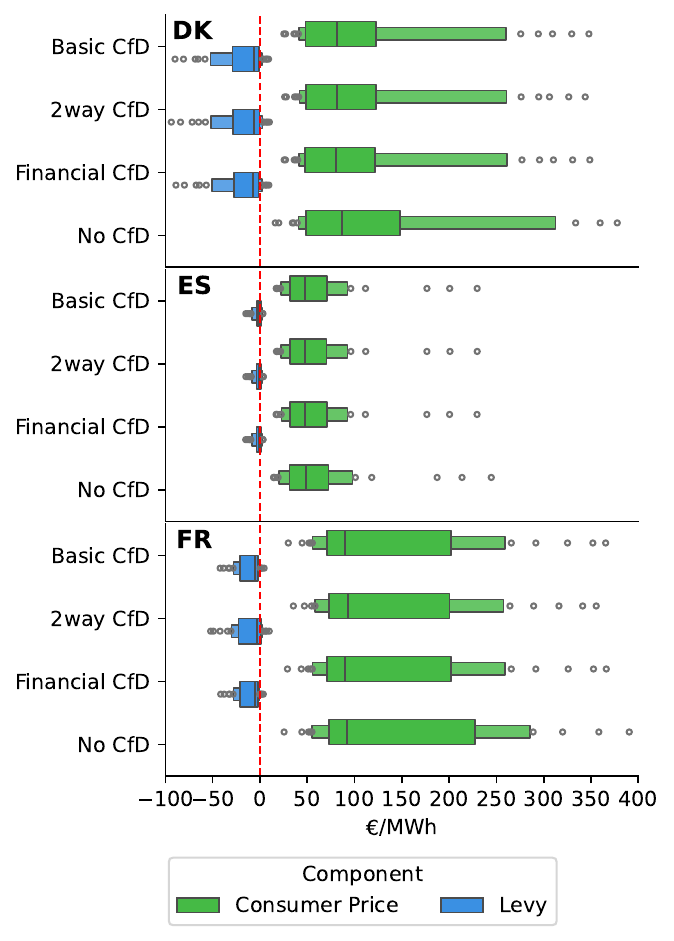}
    \hfill
    \caption{Distribution of consumer price composed of electricity price (not depicted) and levy across 36 dispatch scenarios without a price cap in Denmark (DK), Spain (ES) and France (FR).}
    \label{fig:cons_costs_uncapped}
\end{subfigure}
\caption{Distribution of cost recovery and consumer prices across 36 scenarios without an electricity price cap}
\end{figure}

\newpage
\begin{figure}[!ht]
\begin{subfigure}[t]{0.48\linewidth}
    \includegraphics[width=\columnwidth]{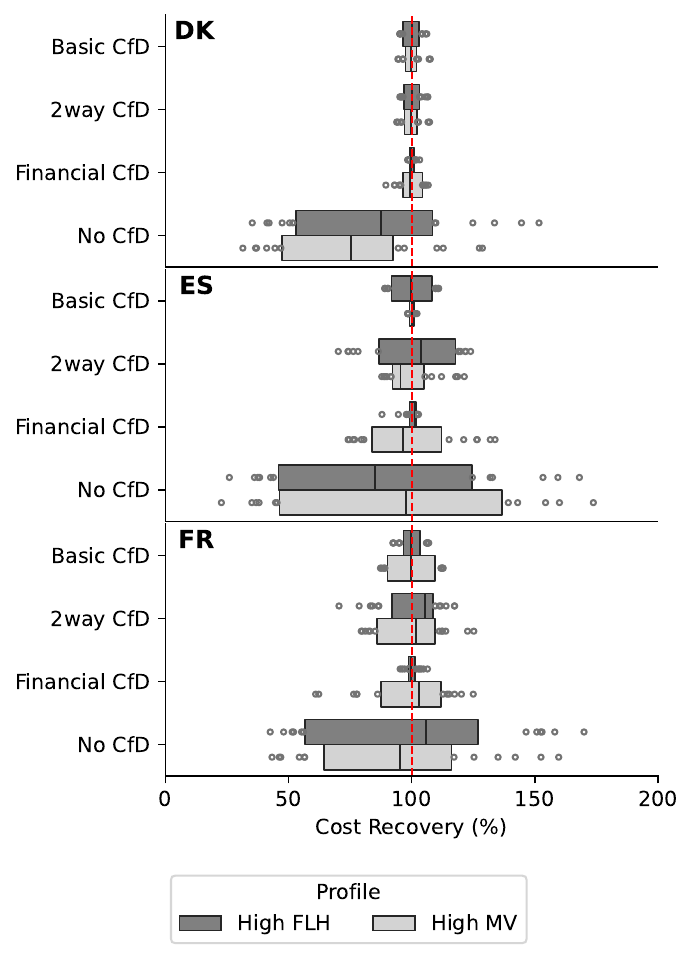}
    \label{fig:ex_post_wo2010}
    \caption{Distribution of ex post cost recovery across 24 dispatch scenarios that only consider the weather years 2015 and 2019 for two wind power plants in Denmark (DK), Spain (ES) and France (FR).}
\end{subfigure}
\hfill
\begin{subfigure}[t]{0.48\linewidth}
    \includegraphics[width=\columnwidth]{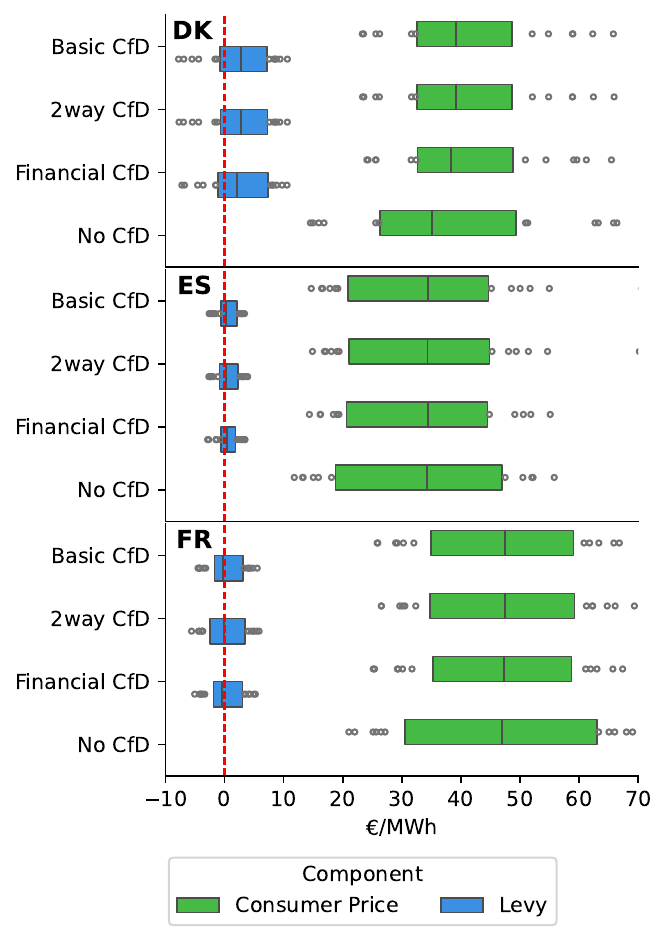}
    \hfill
    \caption{Distribution of consumer price composed of electricity price (not depicted) and levy across 24 dispatch scenarios that only consider the weather years 2015 and 2019 in Denmark (DK), Spain (ES) and France (FR).}
    \label{fig:cons_costs_wo2010}
\end{subfigure}
\caption{Distribution of cost recovery and consumer prices across 24 scenarios without the weather year 2010}
\end{figure}

\end{document}